\documentclass{article}
\usepackage{amsmath}
\usepackage{amsfonts}
\usepackage{amssymb}
\usepackage{graphicx}% Include figure files
\usepackage{dcolumn}% Align table columns on decimal point
\usepackage{bm}% bold mathematics

\title{Trapping in quantum chains}

\author{J.C. Eilbeck${}^{1}$ and  F. Palmero${}^{1,2}$\\
${}^{1}$ Department of Mathematics, 
Heriot-Watt University \\Riccarton, Edinburgh, EH14 4AS, UK\\
${}^{2}$ permanent address: Departamento de F\'{\i}sica Aplicada I.\\ 
ETS Ingenier\'{\i}a Inform\'atica. Universidad de Sevilla\\
Avda Reina Mercedes s/n, 41012-Sevilla, Spain
}

\begin{document}

\maketitle


\begin{abstract}
  A quantum breather on a translationally invariant one-dimensional
  anharmonic lattice is an extended Bloch state with two or more
  particles in a strongly correlated state.  We discuss several
  effects that break the lattice symmetry and lead to spatial
  localization of the breather.

\end{abstract}

\section{Introduction}

The localization of energy by nonlinearity in classical lattices has
been much studied in the last 20 years. The corresponding localized
states, known as intrinsic localized modes or discrete breathers, have
been the subject of intense theoretical and experimental
investigation \cite{Brerev}.  Corresponding results on the
quantum equivalent of discrete classical breathers are less numerous,
c.f.\ \cite{seg94,QTrev} for some theoretical results and \cite{QErev}
for some experimental work.  Studies of quantum modes on small
lattices may be relevant to studies of quantum dots and
quantum computing (c.f.\ \cite{li03}).  

In this paper, we present some results related to quantum lattice
problems, in particular in one dimensional lattices with a small
number of quanta. We study a periodic lattice with $f$ sites
containing bosons, described by the quantum discrete nonlinear
Schr\"{o}dinger equation (QDNLS), a quantum version of the discrete
nonlinear Schr\"{o}dinger equation (also know as the Boson Hubbard
model).  The DNLS equation describes a particularly simple model for a
lattice of coupled anharmonic oscillators, and it has been used to
describe the dynamics of a great variety of systems \cite{sc99}. The
corresponding quantum Hamiltonian is given by
\begin{equation}
\hat H=-\sum_{j=1}^{f} \frac12\gamma_j b_j^\dag b_j^\dag b_j b_j+ 
\epsilon_j b_j^\dag(b_{j-1}+b_{j+1}),
\label{Ham_bos}
\end{equation}
where $b_j^\dag$ and $b_j$ are standard bosonic
operators, $\gamma_j/\epsilon_j$ is the ratio of
anharmonicity to nearest neighbor hopping energy, and the chain is
subject to periodic boundary conditions with period $f$.  Initially we
consider the case where the chain is translationally invariant, i.e.
$\gamma_j=\gamma$ and $\epsilon_j=\epsilon$ are independent of $j$.
In general we take $\epsilon=1$.

The Hamiltonian (\ref{Ham_bos}) has an important conserved quantity,
the number $N=\sum_{j=1}^{f} b_j^\dag b_j$, which enables the total
Hamiltonian to be block-diagonalised and greatly simplifies the
analysis.  In this letter we restrict ourselves to the simplest
nontrivial case $N=2$, though many of the results are valid for larger
values on $N$.  The $N=2$ bound states corresponds to bound two-vibron
states, as observed experimentally in several systems \cite{Jak02}.

In QDNLS case, we use a number state basis,
$|\psi_n\rangle=[n_1,n_2,...,n_f]$, where $n_i$ represents the number
of quanta at site $i$ ($N=\sum n_i$).  A general wave function is
$|\Psi_n\rangle=\sum_n c_n|\psi_n\rangle$.

In homogeneous quantum lattices with periodic boundary conditions, it
is possible to block--diagonalize the Hamiltonian operator using
eigenfunctions of the translation operator with fixed value of the
momentum $k$ \cite{seg94}.

\begin{figure}[h]
\begin{center}
 \includegraphics[scale=0.48]{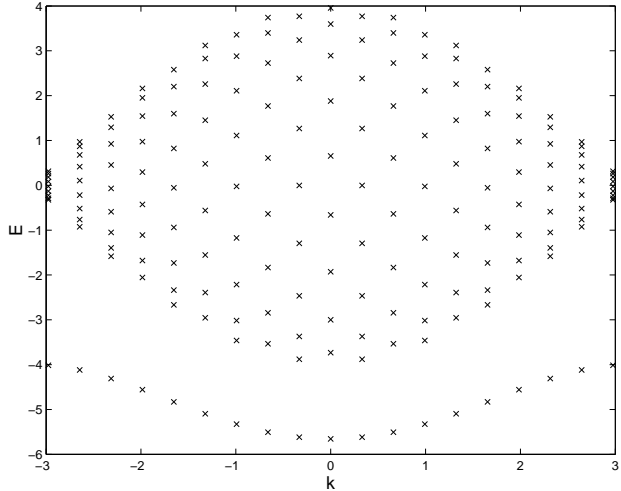}
  \caption{Eigenvalues $E(k)$ for QDNLS model with periodic boundary
 conditions. $N=2$, $f=19$, $\gamma=4$.}
  \label{fig1}
\end{center}
\end{figure}

As shown in Fig.\ \ref{fig1}, if the anharmonicity parameter is high
enough, there exists an isolated eigenvalue for each $k$ which
corresponds to a localized eigenfunction.  By this we mean there is a
high probability of finding the two quanta on the same site, but due to
the translational invariance of the system, an equal probability of
finding these two quanta at any site of the system. In these cases,
some analytical expressions can be obtained in some asymptotic limits
(solutions of this problem go back to the 30's, for recent
discussions see \cite{sc99,seg94,ei02}).

In particular, working at $k=0$ for simplicity, the ground state
unnormalized eigenfunction is
$$
|\Psi\rangle=[20\dots0]+[020\dots0]+\dots+[0\dots02]+O(\gamma^{-1}),
$$
i.e. on a lattice of length $f$, the unnormalized coefficients
$c_i$ of the first $f$ terms are equal to unity and the rest are $O(
\gamma^{-1})$.  In the other extreme of complete spatial
localization, one of these $c_i$ would be unity and the rest zero.  In
this letter we consider how these components change as the
translational invariance of the lattice is broken in various ways.

One simple way that translational invariance can be broken is by
considering a finite chain with no-flux boundary conditions. The
Hamiltonian operator now cannot be block-diagonalized
using eigenvectors of the translation operator. In this case, the
computational effort increases, but it is still possible to calculate
the all the eigenvalues and eigenvectors of the Hamiltonian operator,
if $f$ and $N$ are small enough, by using algebraic manipulation
methods and numerical eigenvalue solvers.

\begin{figure}[h]
  \begin{center}
    \includegraphics[scale=0.25]{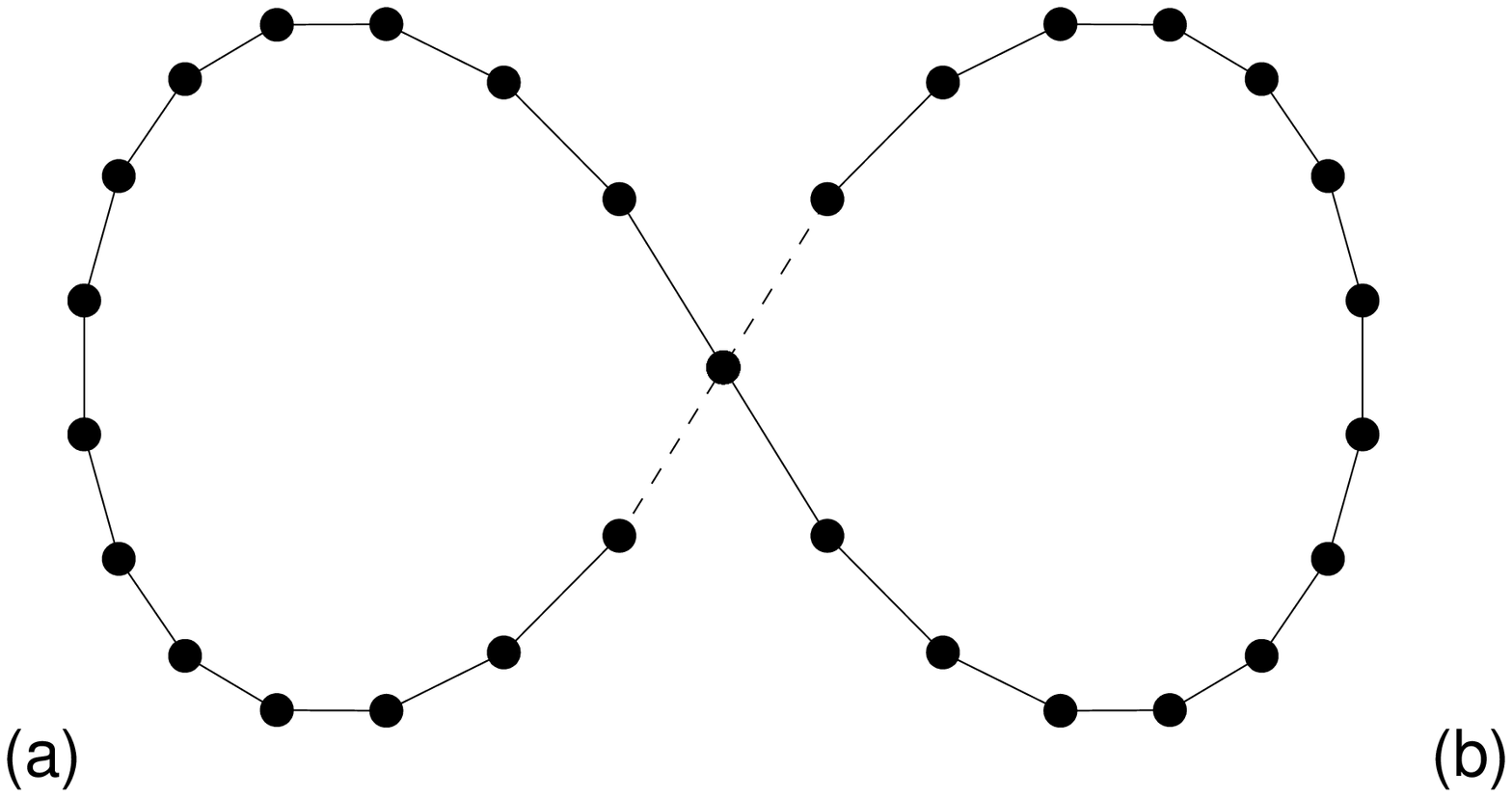}
    \includegraphics[scale=0.3]{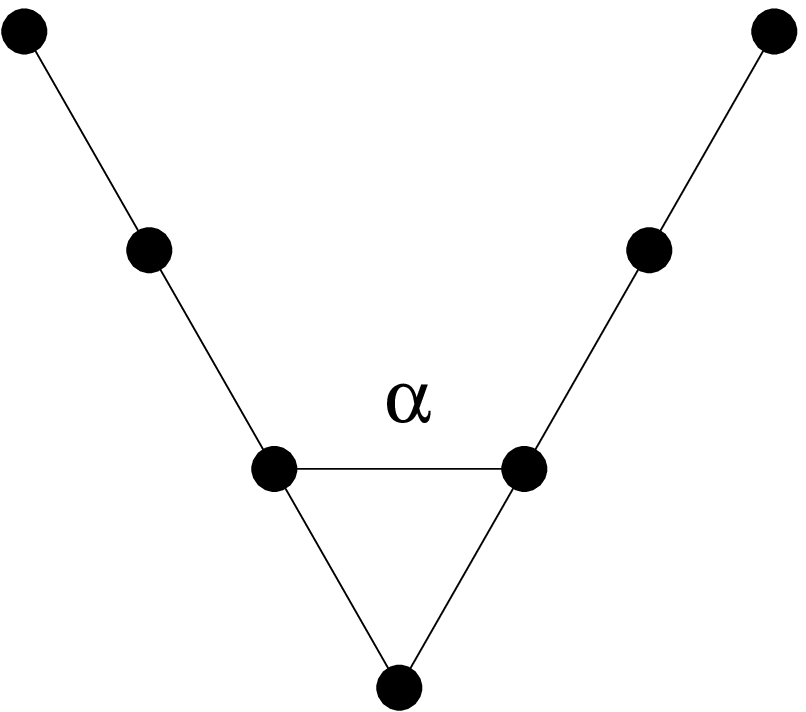}
  \end{center}
  \caption{Two non-uniform chain geometries}
  \label{fig2}
\end{figure}

The existence of local inhomogeneities or impurities in a system can
affect the nonlinear localized modes considerably. Also, in systems
with both nonlinearity and impurities, it is important to understand
the interplay between these two sources of localization.  For these
cases we break translational invariance by making one or more of the
$\gamma_j$ or the $\epsilon_j$ depend on $j$.  This may occur because
of localized impurities, or because the chain geometry becomes
non-uniform.  Two examples of non-uniform geometries are shown in Fig.
\ref{fig2}.

Fig.\ \ref{fig2}a shows a circular chain twisted into a
figure-of-eight, so that two sites on the chain, which are distant
measured along the length of the chain, become spatially close.  This toy
model for a globular protein was studied in \cite{ei86}, where it was
shown that moving breathers described by the classical DNLS equation
could become trapped at the cross-over point.  In the quantum case
such a geometry can be modeled by adding a term such as
\begin{equation}
\alpha_{\ell,m}(b_\ell^\dagger b_m+b_m^\dagger b_\ell), \label{lrterm}
\end{equation}
to the Hamiltonian, where $\ell$ and $m$ are the two sites brought
close together by the twist, and $\alpha_{\ell,m}$ is the separation
distance relative to the unit length of the unperturbed chain.  This
can be considered as a special case of a chain with long-range
coupling.  See also \cite{fe91} for a more realistic protein
simulation, and \cite{cur} for other discussions on the effects of
chain geometry on moving breathers.
 
The bent chain in Fig.\ \ref{fig2}b shows another possible geometry
which has been studied recently in the classical DNLS case. In this
case we have an abrupt bend which is simulated by adding an additional
term as in (\ref{lrterm}) but where $m=m_0-1,\ell =m_0+1$, where $m_0$
is the vertex of the bend.  By varying the values of
$\alpha_{\ell,m}=\alpha$, all angles between 0 and $\pi$ can be simulated
approximately.  The influence of this geometry has been analyzed in
the DNLS context in \cite{Kiv03}, and in nonlinear Klein--Gordon
systems in \cite{Cue03}.  This geometry is of interest in nonlinear
photonic crystals waveguides and circuits \cite{Min02}.

Localization due to random variation of the lattice parameters has long been
studied in the {\it harmonic} model since the pioneering work of
Anderson \cite{an58}.  Our interest is to see what new localization
effects the anharmonic terms bring to the model, and to what extent
the anharmonic effects enhance the Anderson-like localization effect when
this is present in the harmonic model ($\gamma_i=0$).  See also the
discussions in \cite{amm99}.

Most of our findings are not specific to the QDNLS model, for example
we have repeated our calculations using the attractive fermionic
Hubbard model with two particles of opposite spins.  Details will be
given elsewhere.  Although all the models we consider have a conserved
number, in general they are not quantum integrable.

We now consider the effects introduced above in more detail.

\section{Localization in an straight chain with impurities}

In this version of our model, in order to explore the interplay
between the localization induced by the nonlinearity and the influence
of a impurity in these localized states, we introduce a local
inhomogeneity in the anharmonic parameter. To isolate the effect of
the impurity of other effects related to the finite size of the chain,
we retain the periodic boundary conditions. The anharmonicity
parameter is $\gamma_{\ell}=\gamma_{\text{im}}$, and
$\gamma_{j}=\gamma$ for $j \neq \ell$.

\begin{figure}
  \begin{center}
    \includegraphics[scale=0.48]{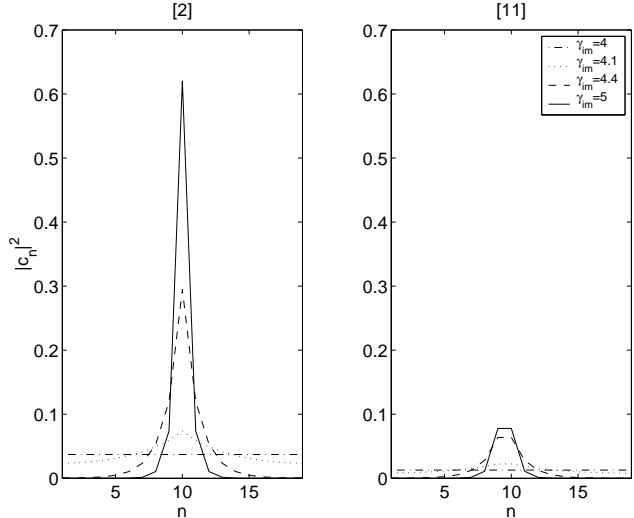}
  \end{center}
  \caption{QDNLS model with point impurity, $\gamma=4$, $N=2$,
    $f=19$.}
  \label{fig3}
\end{figure}

In the homogeneous system ($\gamma_{\text{im}}=\gamma$), with $\gamma$
large enough, as discussed above, the ground state is a localized in
the sense that there exist a high probability to find the two quanta
on the same site, but with equal probability at any site of the chain.
For the chain with a point impurity, we plot in Fig.\ \ref{fig3} the
coefficients of some of the components of the ground state wave
function for various values of $\gamma_\text{im}$.
The left hand figure shows the coefficients of the
$[20\dots],[020\dots],\dots$ components.  As  $\gamma_\text{im}$
increases, these coefficients start from an initial spatially
uniform distribution, but then localize around the site of the
impurity, in this case at $\ell=10$. At the largest value of
$\gamma_\text{im}$ shown, over 60\% of the wave function is in the
state $[0\dots020\dots]$, with the two bosons at site 10.

The right hand figure shows the corresponding coefficients of the
$[110\dots],[0110\dots],\dots$ components.  Again some localization is
found as $\gamma_\text{im}$
increases, but this effect is much weaker for these components.  Other
components are even smaller.

\begin{figure}[h]
  \begin{center}
 \includegraphics[scale=0.48]{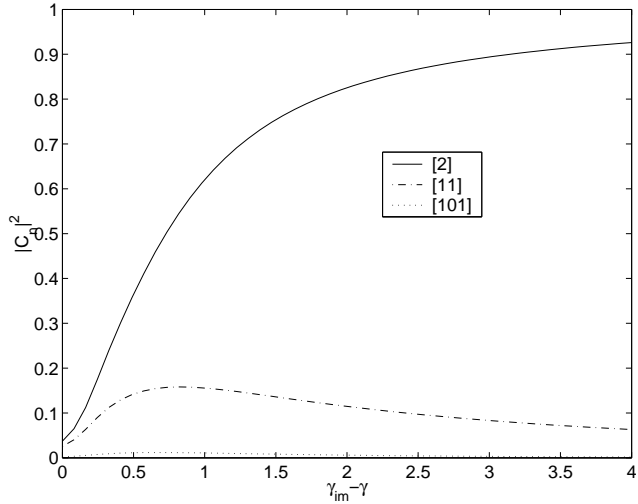}
  \end{center}
  \caption{Some components of the wave function corresponding to 
    the ground state, $N=2, f=19$, $\gamma=4$.}
  \label{fig4}
\end{figure}

In Fig.\ \ref{fig4}, we plot the size of the first three components of
the type $[2], [11], [101]$ respectively, centred around site
10, as a function of $\gamma_\text{im}-\gamma$.  The localization
increases very rapidly with the magnitude of the impurity.
Note that there is no Anderson-like effect in this case as the harmonic
terms are homogeneous.

\section{The twisted chain}
\begin{figure}
  \begin{center}
 \includegraphics[scale=0.48]{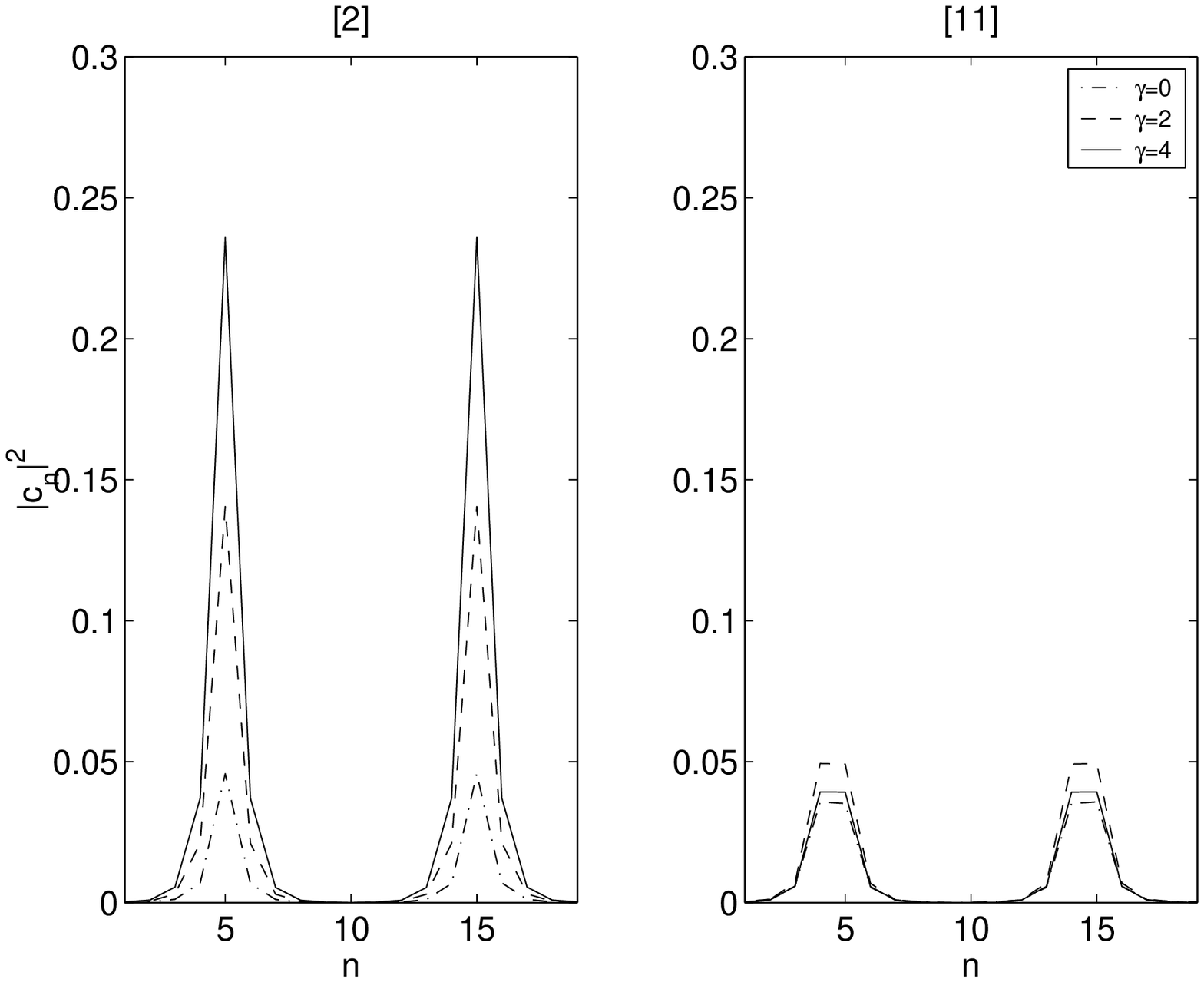}
  \end{center}
  \caption{QDNLS model with twisted chain, $\gamma=4.0$, $N=2$, $f=19$}
  \label{fig5}
\end{figure}

For the twisted chain as shown in Fig.\ \ref{fig2}a, the
only extra parameter is $\alpha_{m,\ell}$, the strength of the long range
coupling between the two spatially adjacent sites $m$ and $\ell$.  As
an example we consider the case 
$\alpha_{m,\ell}=1, f=19, m=5$, and $\ell=15$.   Fig.\ 
\ref{fig5} shows two sets of components of the ground state wave
function, plotted as a function of $\gamma$. 
The components here are the same as those plotted in Fig.\ \ref{fig3},
with a crossover point at $m=5, \ell=15$.  A breather localized
at the crossover point will show an enhanced coefficient corresponding
to localization at the two points of the chain which come together.
With $\gamma=0$, we see some small localization effect at sites $m$
and $\ell$ for the $[2]$ coefficients at the crossover points,
corresponding to a harmonic Anderson-like effect.  However for nonzero
$\gamma$ we see that this effect is strongly enhanced.  The
coefficient of the $[11]$ component show a weaker localization as
before.  Similar results are obtained for other values of
$\alpha_{m,\ell}$, with the strength of the localization depending on
the size of $\alpha_{m,\ell}$.
These trapped quantum states simulate the trapping of the classical
mobile breather studied in \cite{ei86}.

\section{The Bent Chain}

For the bent QDNLS chain Fig.\ \ref{fig2}, we 
follow the classical DNLS treatment \cite{Kiv03}. We consider the
Hamiltonian (\ref{Ham_bos}) on a finite lattice with the additional
term (\ref{lrterm}).  The parameter $\alpha$ is related to the
wedge angle $\theta$ through $\alpha=\frac12(1-\cos \theta)^{-1}$,
and we take the site of the vertex  to be $m_0=\frac12(f+1)$.

\begin{figure}
  \begin{center}
 \includegraphics[scale=0.48]{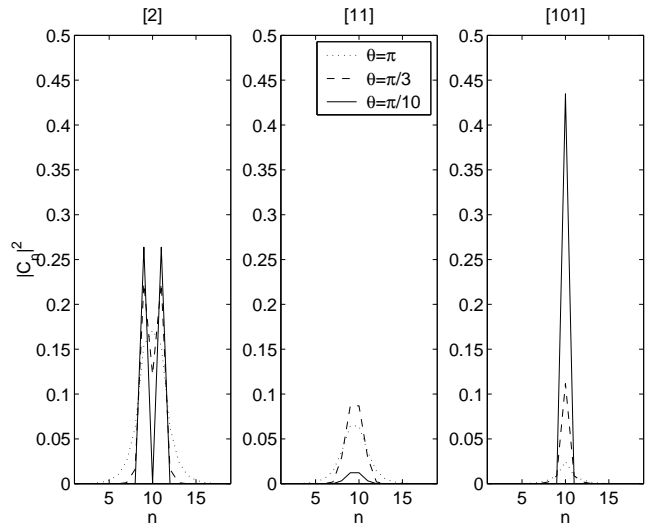}
  \end{center}
   \caption{QDNLS bent chain model, $N=2$, $f=19$,
     $\gamma=4$.}
  \label{fig6}
\end{figure}

\begin{figure}
  \begin{center}
 \includegraphics[scale=0.48]{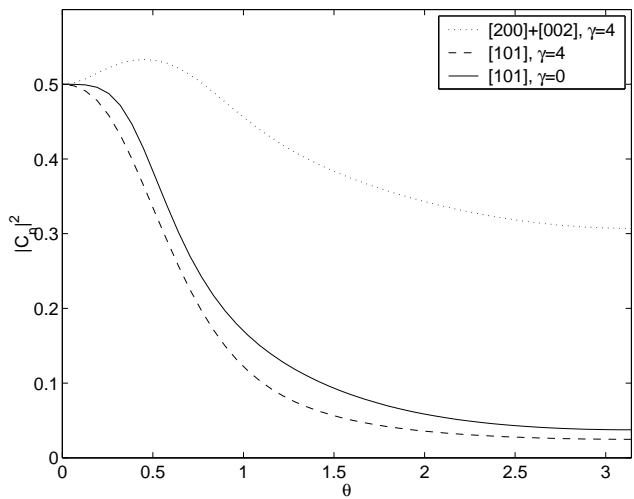}
  \end{center}
  \caption{QNLS bent chain, $N=2, f=19$, various $\gamma$.}
  \label{fig7}
\end{figure}

Fig.\ \ref{fig6} shows the coefficients of the $[2],[11],[101]$
components of the wave function for various values of $\theta$.
Fig.\ \ref{fig7} shows some components of the ground state wave function
corresponding to the neighbors of the vertex, for both 
$\gamma=0$ and  $\gamma=4$. If the angle $\theta$ is close to
$\pi$, the behavior is similar to the straight chain.  The ground
state is weakly localized around the center (vertex) of the chain.  As
$\theta$ decreases, the localization around the vertex increases, and
when this angle is small enough, the largest components of the wave
function in the ground state consists of states localized around the
vertex and the two connected neighboring sites.  In the limit
$\theta\rightarrow 0$, the lattice becomes a T-junction, a model of
interest in its own right.  It is interesting that the localization in
the anharmonic model exhibits a maximum at $\theta \approx 0.5$,
whereas in the harmonic case the maximum is at $\theta=0$.
Also the enhancement due to the anharmonic terms goes to zero as
$\theta \rightarrow 0$.

In this paper, we have considered only a long--range interaction
between the two vertices of the chain.  We have also studied more
realistic models with long--range interaction between
all neighbors in the chain. Qualitatively the same localization
phenomena is observed: a full
description will be given elsewhere.

\section*{Acknowledgements}
The authors are grateful for partial support under the LOCNET EU
network HPRN-CT-1999-00163. F. Palmero thanks Heriot-Watt University
for hospitality, and the Secretar\'{\i}a de Estado de
Educaci\'on y Universidades (Spain) for financial support.

\end{document}